\documentclass[sigconf]{acmart}

\setcopyright{acmcopyright}
\AtBeginDocument{%
  \providecommand\BibTeX{{%
    \normalfont B\kern-0.5em{\scshape i\kern-0.25em b}\kern-0.8em\TeX}}}

\usepackage{pdfpages}
\usepackage{listings}
\usepackage{color}

\definecolor{codegreen}{rgb}{0,0.6,0}
\definecolor{codegray}{rgb}{0.5,0.5,0.5}
\definecolor{codepurple}{rgb}{0.58,0,0.82}
\definecolor{backcolour}{rgb}{0.95,0.95,0.92}
\definecolor{azure}{rgb}{0.0, 0.5, 1.0}
\definecolor{blue(ryb)}{rgb}{0.01, 0.28, 1.0}
\definecolor{blue(pigment)}{rgb}{0.2, 0.2, 0.6}
\setcopyright{rightsretained}

\begin{document}

\copyrightyear{2020} 
\acmYear{2020}
\acmConference[CF '20]{17th ACM International Conference on Computing Frontiers}{May 11--13, 2020}{Catania, Italy}
\acmBooktitle{17th ACM International Conference on Computing Frontiers (CF '20), May 11--13, 2020, Catania, Italy}\acmDOI{10.1145/3387902.3394038}
\acmISBN{978-1-4503-7956-4/20/05}

\title{Enabling Mixed-Precision Quantized Neural Networks in Extreme-Edge Devices}

\author{Nazareno Bruschi}
\affiliation{%
 \institution{University of Bologna, Italy}}
\author{Angelo Garofalo}
\affiliation{%
 \institution{University of Bologna, Italy}}
\author{Francesco Conti}
\affiliation{%
 \institution{University of Bologna, Italy}}
\additionalaffiliation{%
 \institution{ETH Zurich, Switzerland}}
\author{Giuseppe Tagliavini}
\affiliation{%
 \institution{University of Bologna, Italy}}
\author{Davide Rossi}
\affiliation{%
 \institution{University of Bologna, Italy}}

\renewcommand{\shortauthors}{Nazareno Bruschi, Angelo Garofalo, Francesco Conti, Giuseppe Tagliavini and Davide Rossi} 

\begin{abstract}

The deployment of Quantized Neural Networks (QNN) on advanced microcontrollers requires optimized software to exploit digital signal processing (DSP) extensions of modern instruction set architectures (ISA). As such, recent research proposed optimized libraries for QNNs (from 8-bit to 2-bit) such as CMSIS-NN and PULP-NN. This work presents an extension to the PULP-NN library targeting the acceleration of mixed-precision Deep Neural Networks, an emerging paradigm able to significantly shrink the memory footprint of deep neural networks with negligible accuracy loss. The library, composed of 27 kernels, one for each permutation of input feature maps, weights, and output feature maps precision (considering 8-bit, 4-bit and 2-bit), enables efficient inference of QNN on parallel ultra-low-power (PULP) clusters of RISC-V based processors, featuring the RV32IMCXpulpV2 ISA. The proposed solution, benchmarked on an 8-cores GAP-8 PULP cluster, reaches peak performance of 16 MACs/cycle on 8 cores, performing 21$\times$ to 25$\times$ faster than an STM32H7 (powered by an ARM Cortex M7 processor) with 15$\times$ to 21$\times$ better energy efficiency.

\end{abstract}


\maketitle

\section{Introduction}

An increasing amount of Internet-of-Things (IoT) applications acquire data from low-power sensors and transmit it wirelessly after some forms of compression. Machine Learning (ML) algorithms, and in particular Convolutional Neural Networks (CNNs), provide an effective solution for these applications thanks to their capability to squeeze raw sensor data in a much more dense format (e.g., classes or extracted high-level features). As such, a recent trend lies into deploying deep learning functionality on embedded microcontrollers (MCU), which are the de-facto standard compute platform for IoT end-nodes thanks to their flexibility, low-power, and low-cost.

On the other hand, the computing power and memory footprint of MCUs is often not suitable for implementing state-of-the-art models. A recent trend in embedded CNNs to reduce both computational cost and memory footprint of CNNs is quantization~\cite{bib:quantized}\cite{bib:minimum}. This approach, representing the network weights and features with 8-bit or even smaller data types, such as 4-bit or 2-bit, has demonstrated the capability to reduce the memory footprint of state-of-the-art networks~\cite{bib:memory}, with negligible accuracy loss. Optimized software libraries for Quantized Neural Networks (QNNs) have been proposed by the industry by means of CMSIS-NN library~\cite{bib:cmsis}, targeting \mbox{16-bit} and 8-bit QNNs on Cortex-M microcontrollers; as well as by the research community, such as PULP-NN, an open-source library targeting RISC-V processors, and supporting heavily quantized CNNs working on 8-bit, 4-bit, 2-bit, or 1-bit data~\cite{bib:pulpnn}. To further reduce the memory footprint, recent works show how mixed-precision quantization can achieve better performance compared to symmetrical quantization of input feature maps (\textit{ifmaps}), weights, and output feature maps (\textit{ofmaps}). For example, applying this approach on a MobileNetV1 CNN achieves 7$\times$ memory footprint reduction, while incurring an accuracy loss of only 4\%~\cite{bib:mixed} with respect to the 32-bit integer representation.

With this aim, we propose an extension to the PULP-NN open-source library, which includes 27 convolution kernels, one for every permutation of \textit{ifmaps}, weights, and \textit{ofmaps} quantization level, for 8-bit, 4-bit and 2-bit, overtaking the limitations of the current open-source library supporting symmetrical quantization only.
Our solution can reach 16 MACs per cycle on octa-core execution on the GreenWaves Technologies GAP-8~\cite{bib:gap} processor, up to 25$\times$ and 46$\times$ faster with up to 45$\times$ and 21$\times$ less energy than the execution on a STM32H7 and STM32L4, which are commercial MCUs based on an ARM Cortex M7 and M4 core respectively.

\section{Background}

\subsection{QNNs and Mixed-Precision QNNs}
\label{sec:background_qnn}
A QNN is defined by mapping all real-valued tensors involved in a DNN layer (weights $\mathbf{w}$, \textit{ifmaps} $\mathbf{x}$, \textit{ofmaps} $\mathbf{y}$) to integers.
In this work we focus on layer-wise linear quantization, where each real-valued tensor $\mathbf{t}$ in the range $[\alpha_\mathbf{t}, \beta_\mathbf{t})$ is built such that:
\begin{equation}
    \mathbf{t} = \alpha_\mathbf{t} + \varepsilon_\mathbf{t}\cdot{INT}(\mathbf{t}) \label{eq:1}
\end{equation}
where ${INT}(\mathbf{t})$ is an $N$-bit integer-valued tensor with the same dimensionality of $\mathbf{t}$, and $\varepsilon_\mathbf{t} = (\beta_\mathbf{t}-\alpha_\mathbf{t}) / 2^{N}$.
We further constrain $\alpha_\mathbf{x} = \alpha_\mathbf{y} = 0$ for \textit{ifmaps} and \textit{ofmaps}.
A QNN of this kind can be trained efficiently by means of linear quantization-aware training~\cite{bib:pact}, which produces a QNN using real-valued tensors of the form of Eq.~\ref{eq:1}.
The application of \textit{linear} layers (e.g., convolutional and dense), normalization (e.g., batch-norm) and activation (e.g., ReLU) in a QNN can then be mapped to a linear operation combined with a pointwise normalization/activation, working directly on the integer-valued tensors:
\begin{equation}
    {INT}(\mathbf{y}) = \textit{quant}\Big( \textit{linear} \big({INT}(\mathbf{w}), {INT}(\mathbf{x})\big) \Big) \label{eq:2}
\end{equation}
Notice that the accumulator $\varphi \doteq \textit{linear}\big({INT}(\mathbf{w}), {INT}(\mathbf{x})\big)$ is still integer-valued, but requires in general more bits than its inputs (i.e., $\varepsilon_\varphi$ will be smaller than $\varepsilon_\mathbf{x}$ and $\varepsilon_\mathbf{w}$).
\textit{quant} normalizes $\varphi$ with an affine transformation of parameters $\kappa$ and $\lambda$, then collapses its values, ``converting'' it to a representation with less bits (i.e., with bigger $\varepsilon_\mathbf{y}$) \footnotemark:
\begin{equation}
	{INT}(\mathbf{y}) = \textit{quant}(\varphi) = \mathrm{clip}_{[0,\beta_\mathbf{y})} \Big(\Big\lfloor \big(\kappa\cdot{INT}(\varphi)+\lambda\big)\cdot {\varepsilon_\varphi  }/{\mathbf{\varepsilon_\mathbf{y}}}\Big\rfloor\Big)
\end{equation}
\footnotetext{The $\kappa$ and $\lambda$ parameters can be integrated directly in the ladder function, resulting in a \textit{quant} function that produces ${INT}(\mathbf{y})$ by comparing $\varphi$ with a set of $2^N$ \textit{thresholds}~\cite{bib:memory}.}

In mixed-precision QNNs, the number of bits used for $\mathbf{w}$, $\mathbf{x}$ and $\mathbf{y}$ is not constrained to be the same.
This class of QNNs have been shown~\cite{bib:mixed} to better fit embedded constraints while incurring in a less severe accuracy hit than non-mixed-precision QNNs; massive memory gains can be exploited on tensors and layers that are less sensitive to strong quantization while still keeping more sensitive ones at a higher precision.
Here, we focus on 2-, 4- and 8-bit quantization for $\mathbf{w}$~(signed), $\mathbf{x}$ and $\mathbf{y}$~(unsigned), while we always consider 32 bits for the accumulator $\varphi$ (signed).

\subsection{PULP-NN}\label{back_B}

The software solution we propose is built upon an open-source Parallel Ultra-Low-Power (PULP) cluster of eight RISC-V based processors \footnote{https://github.com/pulp-platform/pulp}. The cores feature a 4-stage in-order pipeline and the RV32IMC ISA, extended with efficient digital signal processing custom instructions, namely XpulpV2. A detailed description of these extensions can be found in~\cite{bib:xpulp}. The key elements of a PULP cluster are a low-latency multi-banked Tightly Coupled Data Memory (TCDM), enabling shared-memory parallel programming models such as OpenMP or OpenCL, and an Event Unit which manages synchronization and thread dispatching, enabling low-overhead and fine-grained parallelism, guaranteeing high efficiency for parallel workloads. In this work, we leverage a commercial embodiment of the PULP architecture fabricated in CMOS 55nm called GAP-8~\cite{bib:gap}.

\begin{figure}[t]
\centerline{\includegraphics[width=0.9\linewidth]{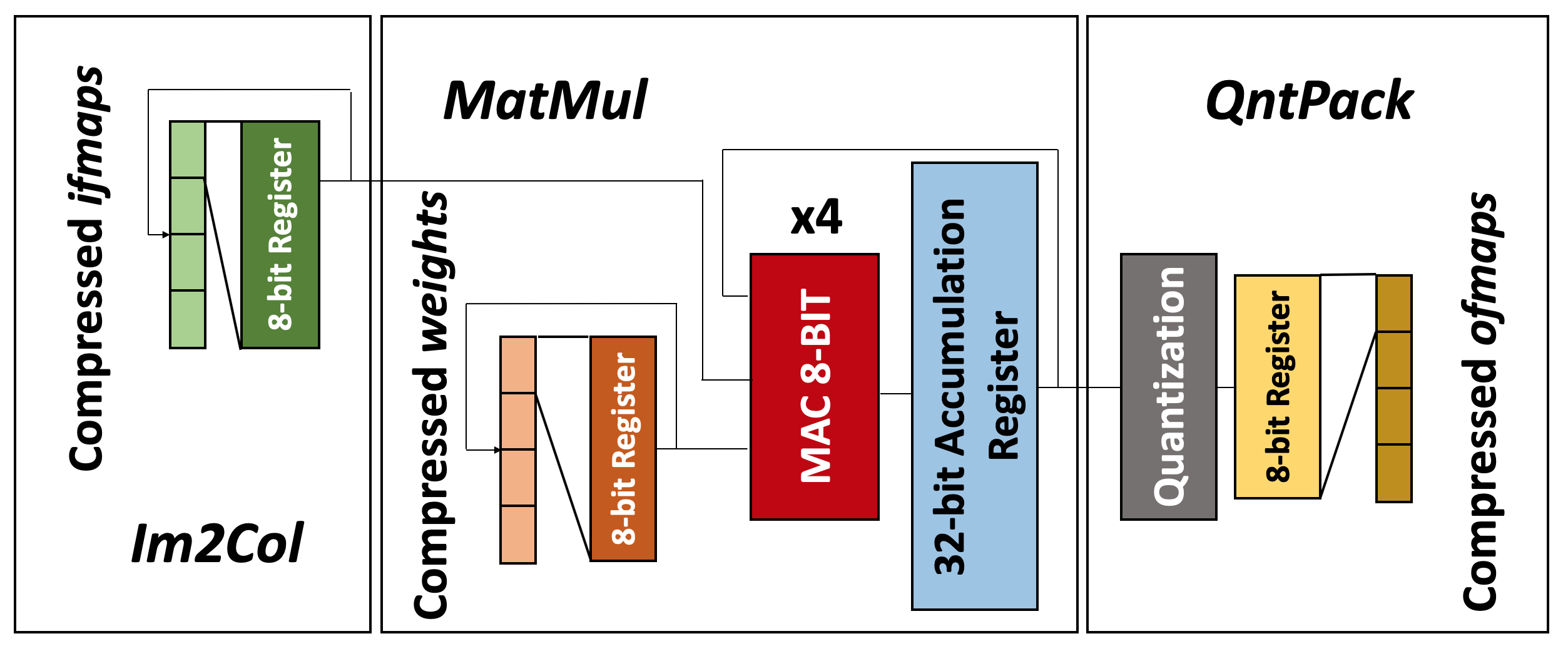}}
\vspace{-2mm}
\caption{Concept scheme of Mixed-Precision Approach.}
\vspace{-1mm}
\label{fig:mixed}
\end{figure}

The PULP-NN library, extended in this work to support mixed-precision QNNs, relies on the Height-Width-Channel (HWC) data layout and on an execution flow optimized to target resource-constrained MCUs. A layer is run as a combination of three phases: the \textit{im2col} step re-arranges the 3D input features of the current layer into a 1D vector, the \textit{linear} part of the layer is implemented through a Matrix Multiplication (\textit{MatMul}) kernel, while a final stage of quantization, named \textit{QntPack}, implements the \textit{quant} function of Eq.~\ref{eq:2}, compresses the \textit{MatMul} result into the desired precision and then stores back the \textit{ofmap}.

The \textit{MatMul} loads the quantized weights from four different filter banks and the input features from two different \textit{im2col} buffers into the register file from the TCDM. Exploiting the data locality of the loaded features and weights within the register file enables computing two spatially adjacent output features of four consecutive output channels in each run of \textit{MatMul} inner loop, optimizing the execution of the kernel. Further details can be found in~\cite{bib:pulpnn}. Since the results of \textit{MatMul} need to be accumulated into higher precision variables, they need to be compressed back to the desired output precision. While for 8-bit output features scaling and clamp operations can be used~\cite{bib:cmsis}, an effective solution for sub-byte outputs consists of a thresholding-based procedure~\cite{bib:mixed, bib:quantized}. This operation compares an input with a set of thresholds (see Section~\ref{sec:background_qnn}). Every \textit{Conv} kernel  presented in this work is parallelized on the H-spatial dimension of \textit{ofmaps}~\cite{bib:pulpnn}, resulting into an almost ideal speed-up on an 8-cores cluster (7.5$\times$).

\section{Mixed-Precision Kernels}\label{descr}

In this section, we describe the proposed mixed-precision software kernels, highlighting the optimizations made to boost the kernels on the PULP cluster.
In the context of a mixed-precision convolution kernel (\textit{Conv}), the precision of the \textit{ifmaps} determines the specific \textit{im2col} function to be used, the precision of the \textit{weights} determines the specific \textit{MatMul} kernel, while the \textit{ofmap} determines the specific \textit{QntPack} kernel. 

\begin{figure}
    \begin{lstlisting}[basicstyle=\tiny]
|\color{magenta}\textbf{void}| pulp_nn_int4_to_int8(int8_t *Src,int8_t *Out)
{
    int8_t bext1 = bext(Src, |\color{codepurple}\textbf{4}|, |\color{codepurple}\textbf{0}| );
    int8_t bext2 = bext(Src, |\color{codepurple}\textbf{4}|, |\color{codepurple}\textbf{4}| );
    int8_t bext3 = bext(Src, |\color{codepurple}\textbf{4}|, |\color{codepurple}\textbf{8}| );
    int8_t bext4 = bext(Src, |\color{codepurple}\textbf{4}|, |\color{codepurple}\textbf{12}|);
    *((v4s*)Out) = pack(bext1, bext2, bext3, bext4);
    Out++;
    bext1 = bext(Src, |\color{codepurple}\textbf{4}|, |\color{codepurple}\textbf{16}|);
    bext2 = bext(Src, |\color{codepurple}\textbf{4}|, |\color{codepurple}\textbf{20}|);
    bext3 = bext(Src, |\color{codepurple}\textbf{4}|, |\color{codepurple}\textbf{24}|);
    bext4 = bext(Src, |\color{codepurple}\textbf{4}|, |\color{codepurple}\textbf{28}|);
    *((v4s*)Out) = pack(bext1, bext2, bext3, bext4);
}
\end{lstlisting}
\vspace{-2mm}
\caption{ Example of code of efficient bit extraction using XpulpV2 extension.}
\vspace{-2mm}
    \label{fig:Ext}
\end{figure}

As the underlying hardware offers support only for 8-bit SIMD instructions, when sub-byte input features are considered, additional unpacking functions must be added to the \textit{im2col} procedure to extract and sign-extend the sub-byte operands into INT-8, natively supported by the sum-of-dot product units. Fig.~\ref{fig:mixed} highlights a general scheme of mixed-precision \textit{Conv} structure. Depending on the \textit{ifmap} and weights precision, different specific casting functions are built, to reduce the number of load instructions needed to fetch the compressed features, hence minimizing the memory traffic and improving the performance. The casting operation takes place also in the innermost loop of the \textit{MatMul} kernel to `unpack` the weight elements. To reduce unpacking operation overhead, we exploit the bit manipulation instructions provided by the XpulpV2 ISA. 

As depicted in Fig.~\ref{fig:Ext} for the 4-bit case without loss of generality, we exploit the XpulpV2 \textit{bit extraction} operation, inferred in the C code as a built-in function (\textit{bext}), which extracts a specified number of bits from a 32-bit register in one clock cycle, also extending the sign bit. With one 32-bit load instruction that fetches eight lower-precision operands, we can obtain eight 8-bit operands (with sign extension), packed in two 32-bit vector registers, ready to be handled in a SIMD fashion. For the 2-bit case, the cost of the load operation is further amortized, since with one 32-bit load we obtain 16 8-bit operands, achieving 0.0625 loads per operand, half than in the 4-bit case.

Since unpacking is a critical operation for \textit{MatMul} kernels, several solutions have been explored, leading to the following optimal kernels structure:

\begin{itemize}
\item 8-bit weights: 6 32-bit loads and 8 SIMD MACs for a total of 14 cycles per iteration.
\item 4-bit weights: 8 32-bit loads, 32 extractions, 16 pack and 16 SIMD MAC, for a total of 72 cycles per iteration.
\item 2-bit weights: like the previous one but, due to the different level of unrolling in the extraction function, there is a total of 140 cycles per iteration.
\end{itemize}

\noindent Each inner loop is iterated over the \textit{im2col} size. The number of iterations depends on the overall number of MACs for each iteration. 

\begin{figure}[t]

\begin{lstlisting}[basicstyle=\tiny]
|\color{magenta}\textbf{void}| pulp_nn_insert_int4(int8_t Src1, int8_t Src2, int8_t *Out)
{
    int8_t mask =  |\color{codepurple}\textbf{0xf0}|;
    int8_t n_mask = not(mask);
    int8_t off = |\color{codepurple}\textbf{0x04}|;
    *Out = bins(Src1, n_mask, Src2, mask, off);
}
\end{lstlisting}
\vspace{-2mm}
\caption{Example of code of efficient bit compression using XpulpV2 extension.}
\vspace{-2mm}
\label{fig:Bitins}   
\end{figure}

Since the \textit{MatMul} works on 32-bit accumulators as specified in Section~\ref{sec:background_qnn}, the \textit{QntPack} function quantizes and pack it to the desired \textit{ofmap} precision. While simple shifts and clamps are used to restore the output range in 8-bit and store it in an 8-bit variable, to compress back sub-byte results in an 8-bit one, additional \textit{packing} functions have to be added after the thresholding-based quantization. This is implemented efficiently, exploiting the \textit{bit insert} function that acts as a natural counterpart of the \textit{bit extract}, which compresses the data and packs them into 8-bit variables (see Fig.~\ref{fig:Bitins}).

\section{Experimental Results}

We ran our kernels on GAP-8 as a commercial product, an edge low-power and octa-core PULP device optimized to perform DNN algorithms~\cite{bib:gap} and then we compared the execution performance and the energy consumption of an STM32H7 and STM32L4 MCU, which run the same layer and the same kernels.
Although the proposed library is fully flexible, we present the results of a reference layer featuring 32x16x16 \textit{ifmaps} size, 64x16x16 \textit{ofmaps} size and 3x3 filters. This layer has a 288 im2col buffer size, referred to in the following as \textit{Reference Layer}, and it is among the ones featuring the best performance on the targeted architectures. We considered the MACs per cycle and cycles per output pixel as the key metrics to define the performance of the library.

\subsection{Single- and Multi-Core Execution Results}

\begin{figure}[t!]
\centerline{\includegraphics[width=0.8\linewidth]{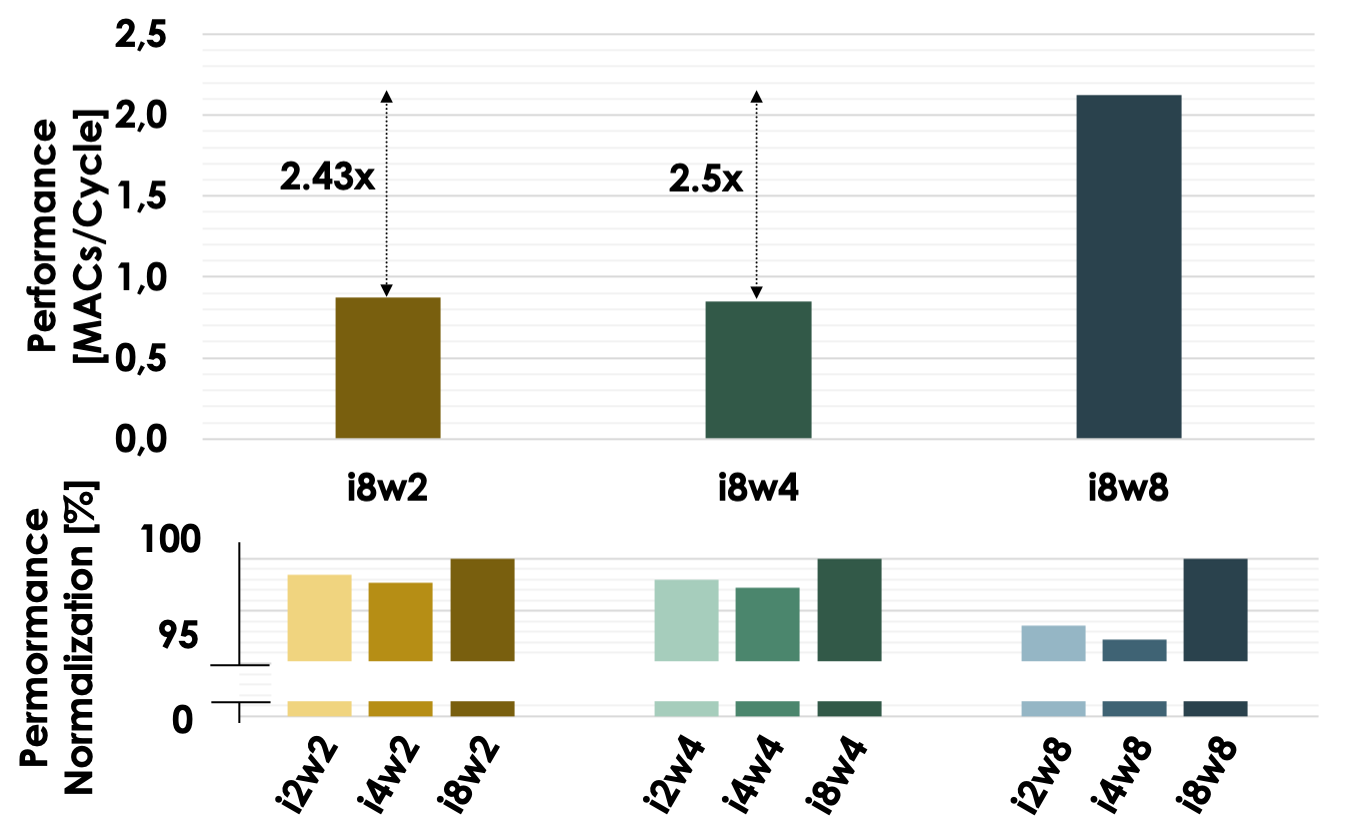}}
\vspace{-2mm}
\caption{MACs/cycle in a single-core \textit{linear} execution in different \textit{weights} precision and its fluctuations by varying \textit{ifmaps} precision.}
\vspace{-1mm}
\label{fig:mac-cyc}
\end{figure}

As seen in Section~\ref{descr}, the inner loop has a different number of iterations  depending on the weights precision level. Therefore, we should expect a decrease in terms of performance of 2.57$\times$ and 2.5$\times$ with respect to the 8-bit \textit{MatMul}, for 4-bit and 2-bit \textit{MatMul}, respectively, due to the unpacking overhead.

To isolate the contributions of the \textit{linear} kernel execution, in Fig.~\ref{fig:mac-cyc} we consider \textit{im2col} and \textit{MatMul} in isolation, removing the per-output-pixel overhead of the \textit{QntPack} function.
The plot shows how much weight unpacking impacts the MACs per cycle performance metric compared to the 8-bit case. In line with expectations, performance drops by 2.43$\times$ and 2.5$\times$ in 2-bit and 4-bit scenarios, respectively. The solution proposed for 2-bit weights is more efficient than 4-bit because it reduces the number of load instructions per MAC, despite introducing more unpacking and packing instructions in the inner loop. Under the bars, Fig.~\ref{fig:mac-cyc} shows how much the \textit{ifmaps} precision impacts performance at a fixed \textit{weights} precision. We observe how the pattern is similar to the one for weights (the solution for 8-bit is the best one, while 2-bit is better than 4-bit); however, it is important to note that the variation is much smaller than that observed when changing weight precision.

\begin{table}[t!]
  \caption{Average overhead cost in cycles per output pixel and its variance by varying the \textit{ofmaps} precision.}
  \vspace{-2mm}
  \label{tab:over}
  \begin{tabular}{ccl}
    \toprule
    \textit{ofmaps} precision&cycles/output pixel&variance\\
    \midrule
    8-bit & 2.01 & +/- 0.57\\
    4-bit & 16.64 & +/- 4.47\\
    2-bit & 8.02 & +/- 1.15\\
  \bottomrule
\end{tabular}
  \vspace{-4mm}
\end{table}

In Tab.~\ref{tab:over}, we investigate the overhead introduced by the \textit{QntPack} function to the overall layer computation.
Due to deep compiler optimization and instruction cache effects, these results have a high variance, which we explicitly represent in the table.
In particular, depending on the size and structure of the innermost loop, code integrating the \textit{linear} and \textit{QntPack} functions is optimized differently in each case, resulting in a different number of inner instructions but also in different binary code sizes, triggering more instruction cache misses in some cases with respect to others.
Despite this significant variability, we can observe clear trends in the overhead \textit{QntPack} introduces.
When using thresholds as activation functions, the average results in Tab.~\ref{tab:over} is as expected, because they are realized with if-else nested statements that perform the binary search in the range in which the output value is found. 4-bit quantization requires twice the number of threshold comparisons than 2-bit quantization, therefore we expect double of cycles per output pixel.
Most non-idealities come from this operation, which is expensive in terms of branches and pipeline stalls.
Furthermore, when we have sub-byte quantization of output, bit compression instructions also enter into the game to pack two or four pixels into an \textit{ofmap} byte, and 8-bit \textit{ofmaps} perform better than 4-bit and 2-bit operations.

\begin{figure}[t!]
\centerline{\includegraphics[width=0.8\linewidth]{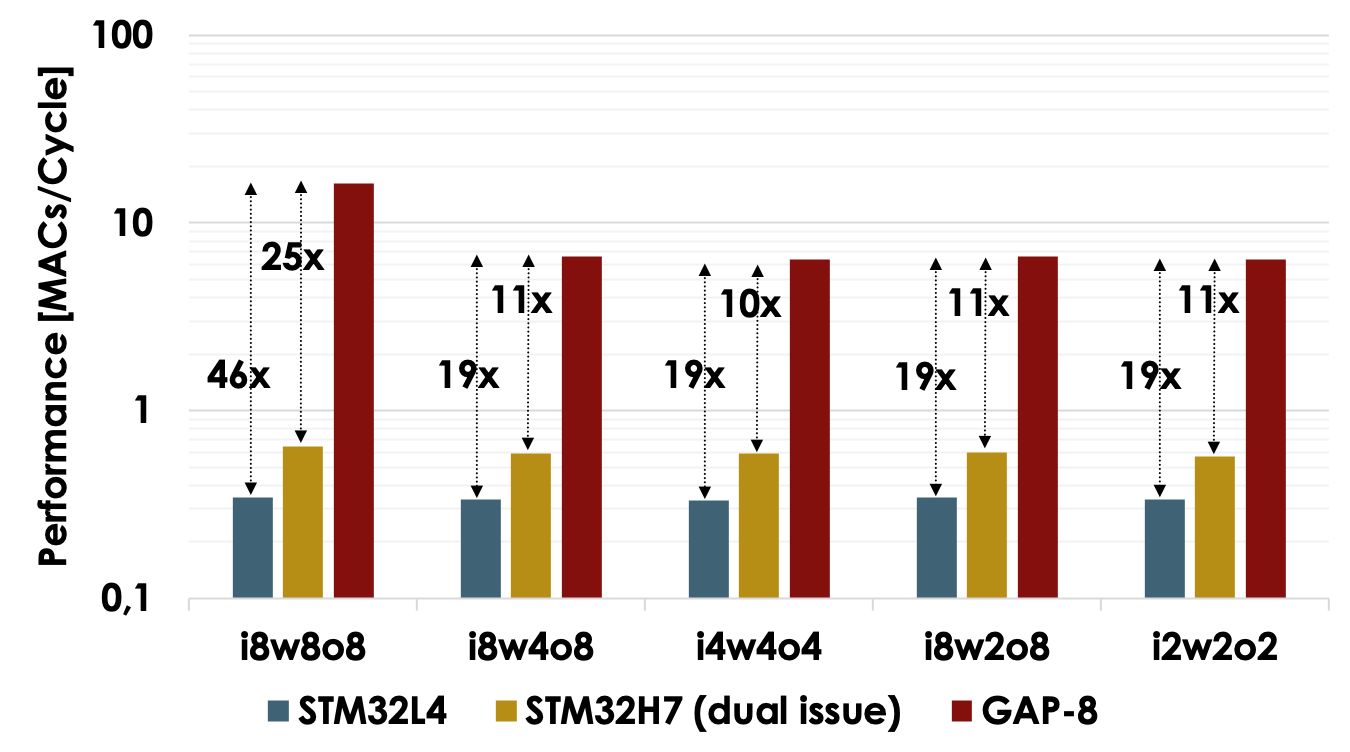}}
\vspace{-2mm}
\caption{Speed-up of PULP-NN Mixed-Precision on GAP-8 (8 cores) over STM32H7 and STM32L4 on the \textit{Reference Layer}.}
\label{fig:h7vsg8}
\end{figure}

\subsection{Comparison with state-of-the-art}

To compare our work with the state-of-the-art, we used an STM32H7 and an STM32L4 running the \textit{Reference Layer}, which are a dual and single issue processors respectively. In Fig.~\ref{fig:h7vsg8} we can see the cycle/cycle speed-up that an octa-core GAP-8 can achieve respect to these devices. In terms of MACs per cycle, we achieve up to 25$\times$ and 46$\times$ in a \textit{Conv} kernel with 8-bit \textit{ifmaps}/\textit{ofmaps} and weights. The contributions of this improvement are certainly to attribute not only to octa-cores execution but also to the XpulpV2 ISA that, compared to ARM Cortex-M7 and -4, features extensions to perform SIMD 8-bit MACs in one cycle with respect to 16-bit SIMD MACs of ARM Cortex-M-based ones. On the other hand, also when unpacking is necessary, we still perform up to 11$\times$ and 19$\times$ respectively.

\begin{figure}[!t]
\centerline{\includegraphics[width=0.8\linewidth]{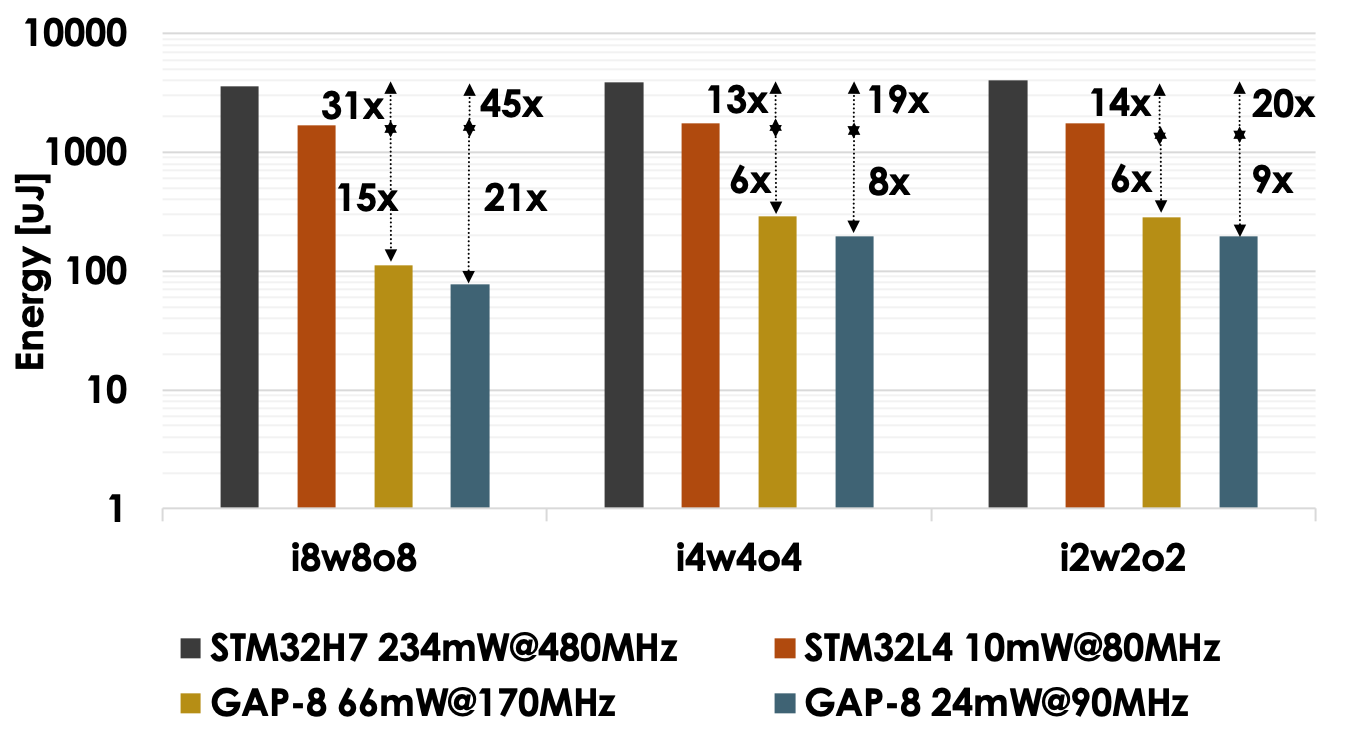}}
\vspace{-2mm}
\caption{GAP-8, STM32H7 and STM32L4 energy consumption when they run the \textit{Reference Layer}.}
\label{fig:energy}
\end{figure}

Finally, we compared the energy consumption of the \textit{Reference Layer} on the benchmarked platforms.
We used the two different operating modes for GAP-8: low-power and high-performance (see Fig.~\ref{fig:energy}).
Despite the less scaled technology node used for the implementation of GAP-8 with respect to STM32H7 (i.e., 55 vs. 40 nm CMOS) and the higher frequency respect to STM32L4 (i.e., 90 MHz vs 80 MHz), GAP-8 performs with 45$\times$ and 21$\times$ less energy consumption in the low-power mode and 31$\times$ and 15$\times$ in the high-performance one, with 8-bit precision operands.
When the unpacking is necessary, the energy consumption stills up to 20$\times$ and 9$\times$ in low-power mode and 14$\times$ and 6$\times$ in high-performance one respect to STM32H7 and STM32L4 execution respectively, as depicted in Fig.~\ref{fig:energy}. This demonstrates the potential of the parallel execution on an optimized cluster with PULP-specific instruction set extensions, coupled with an optimized software abstraction layer able to efficiently exploit the underlying hardware.

\section{Conclusion}

We have presented an open-source software library for mixed-precision inference on parallel ultra-low-power clusters at the edge of the IoT. The proposed library supports 8-bit, 4-bit and 2-bit QNN kernels and for all variants of input feature maps, weights, and output feature maps. Exploiting the DSP capabilities of the XpulpV2 extensions, coupled with the performance gain of parallel execution, our solution can reach up to 16 MACs per cycle on quantized convolutional kernels on the 8-core PULP cluster of the GAP-8 SoC. These results outperform by 25$\times$ the execution of the same kernels on an STM32H7 microcontroller, with 21$\times$ better energy efficiency compared to STM32L4 microcontroller.

\begin{acks}
This work was supported in part by the OPRECOMP project (g.a. 732631) and by the WiPLASH project (g.a. 863337) founded from the European Union’s Horizon 2020 research and innovation program.
\end{acks}

\bibliographystyle{ACM-Reference-Format}
\bibliography{main}


\begin{thebibliography}{9}


\ifx \showCODEN    \undefined \def \showCODEN     #1{\unskip}     \fi
\ifx \showDOI      \undefined \def \showDOI       #1{#1}\fi
\ifx \showISBNx    \undefined \def \showISBNx     #1{\unskip}     \fi
\ifx \showISBNxiii \undefined \def \showISBNxiii  #1{\unskip}     \fi
\ifx \showISSN     \undefined \def \showISSN      #1{\unskip}     \fi
\ifx \showLCCN     \undefined \def \showLCCN      #1{\unskip}     \fi
\ifx \shownote     \undefined \def \shownote      #1{#1}          \fi
\ifx \showarticletitle \undefined \def \showarticletitle #1{#1}   \fi
\ifx \showURL      \undefined \def \showURL       {\relax}        \fi
\providecommand\bibfield[2]{#2}
\providecommand\bibinfo[2]{#2}
\providecommand\natexlab[1]{#1}
\providecommand\showeprint[2][]{arXiv:#2}

\bibitem[\protect\citeauthoryear{et. al.}{et. al.}{2020}]%
        {bib:mixed}
\bibfield{author}{\bibinfo{person}{Alessandro~Capotondi et. al.}}
  \bibinfo{year}{2020}\natexlab{}.
\newblock \showarticletitle{CMix-NN: Mixed Low-Precision CNN Library for
  Memory-Constrained Edge Devices}.
\newblock \bibinfo{journal}{\emph{IEEE Transactions on Circuits and Systems II:
  Express Briefs}} (\bibinfo{year}{2020}), \bibinfo{pages}{1--1}.
\newblock


\bibitem[\protect\citeauthoryear{et. al.}{et. al.}{2019}]%
        {bib:pulpnn}
\bibfield{author}{\bibinfo{person}{Angelo~Garofalo et. al.}}
  \bibinfo{year}{2019}\natexlab{}.
\newblock \showarticletitle{PULP-NN: accelerating quantized neural networks on
  parallel ultra-low-power RISC-V processors}.
\newblock \bibinfo{journal}{\emph{Philosophical Transactions of the Royal
  Society A: Mathematical, Physical and Engineering Sciences}}
  \bibinfo{volume}{378}, \bibinfo{number}{2164} (\bibinfo{date}{Dec}
  \bibinfo{year}{2019}), \bibinfo{pages}{20190155}.
\newblock
\showISSN{1471-2962}
\urldef\tempurl%
\url{https://doi.org/10.1098/rsta.2019.0155}
\showDOI{\tempurl}


\bibitem[\protect\citeauthoryear{et. al.}{et. al.}{2017}]%
        {bib:minimum}
\bibfield{author}{\bibinfo{person}{Bert~Moons et. al.}}
  \bibinfo{year}{2017}\natexlab{}.
\newblock \bibinfo{title}{Minimum Energy Quantized Neural Networks}.
\newblock
\newblock
\showeprint[arxiv]{cs.NE/1711.00215}


\bibitem[\protect\citeauthoryear{et. al.}{et. al.}{2018a}]%
        {bib:gap}
\bibfield{author}{\bibinfo{person}{Eric~Flamand et. al.}}
  \bibinfo{year}{2018}\natexlab{a}.
\newblock \showarticletitle{GAP-8: A RISC-V SoC for AI at the Edge of the IoT}.
  In \bibinfo{booktitle}{\emph{2018 IEEE 29th International Conference on
  Application-specific Systems, Architectures and Processors (ASAP)}}.
  \bibinfo{pages}{1--4}.
\newblock


\bibitem[\protect\citeauthoryear{et. al.}{et. al.}{2016a}]%
        {bib:quantized}
\bibfield{author}{\bibinfo{person}{Itay~Hubara et. al.}}
  \bibinfo{year}{2016}\natexlab{a}.
\newblock \bibinfo{title}{Quantized Neural Networks: Training Neural Networks
  with Low Precision Weights and Activations}.
\newblock
\newblock
\showeprint[arxiv]{cs.NE/1609.07061}


\bibitem[\protect\citeauthoryear{et. al.}{et. al.}{2018b}]%
        {bib:pact}
\bibfield{author}{\bibinfo{person}{Jungwook~Choi et. al.}}
  \bibinfo{year}{2018}\natexlab{b}.
\newblock \bibinfo{title}{PACT: Parameterized Clipping Activation for Quantized
  Neural Networks}.
\newblock
\newblock
\showeprint[arxiv]{cs.CV/1805.06085}


\bibitem[\protect\citeauthoryear{et. al.}{et. al.}{2018c}]%
        {bib:cmsis}
\bibfield{author}{\bibinfo{person}{Liangzhen~Lai et. al.}}
  \bibinfo{year}{2018}\natexlab{c}.
\newblock \bibinfo{title}{CMSIS-NN: Efficient Neural Network Kernels for Arm
  Cortex-M CPUs}.
\newblock
\newblock
\showeprint[arxiv]{cs.NE/1801.06601}


\bibitem[\protect\citeauthoryear{et. al.}{et. al.}{2016b}]%
        {bib:xpulp}
\bibfield{author}{\bibinfo{person}{Michael~Gautschi et. al.}}
  \bibinfo{year}{2016}\natexlab{b}.
\newblock \bibinfo{title}{A near-threshold RISC-V core with DSP extensions for
  scalable IoT Endpoint Devices}.
\newblock
\newblock
\showeprint[arxiv]{cs.AR/1608.08376}


\bibitem[\protect\citeauthoryear{et. al.}{et. al.}{2018d}]%
        {bib:memory}
\bibfield{author}{\bibinfo{person}{Manuele~Rusci et. al.}}
  \bibinfo{year}{2018}\natexlab{d}.
\newblock \showarticletitle{Quantized NNs as the Definitive Solution for
  Inference on Low-Power ARM MCUs? Work-in-Progress}. In
  \bibinfo{booktitle}{\emph{Proceedings of the International Conference on
  Hardware/Software Codesign and System Synthesis}} (Turin, Italy)
  \emph{(\bibinfo{series}{CODES ’18})}. \bibinfo{publisher}{IEEE Press},
  Article \bibinfo{articleno}{Article 12}, \bibinfo{numpages}{2}~pages.
\newblock
\showISBNx{9781538655627}


\end{thebibliography}

\end{document}